

\input harvmac.tex


%
\def\ubrackfill#1{$\mathsurround=0pt
	\kern2.5pt\vrule depth#1\leaders\hrule\hfill\vrule depth#1\kern2.5pt$}
\def\contract#1{\mathop{\vbox{\ialign{##\crcr\noalign{\kern3pt}
	\ubrackfill{3pt}\crcr\noalign{\kern3pt\nointerlineskip}
	$\hfil\displaystyle{#1}\hfil$\crcr}}}\limits
}

\def\ubrack#1{$\mathsurround=0pt
	\vrule depth#1\leaders\hrule\hfill\vrule depth#1$}
\def\dbrack#1{$\mathsurround=0pt
	\vrule height#1\leaders\hrule\hfill\vrule height#1$}
\def\ucontract#1#2{\mathop{\vbox{\ialign{##\crcr\noalign{\kern 4pt}
	\ubrack{#2}\crcr\noalign{\kern 4pt\nointerlineskip}
	$\hskip #1\relax$\crcr}}}\limits
}
\def\dcontract#1#2{\mathop{\vbox{\ialign{##\crcr
	$\hskip #1\relax$\crcr\noalign{\kern0pt}
	\dbrack{#2}\crcr\noalign{\kern0pt\nointerlineskip}
	}}}\limits
}

\def\ucont#1#2#3{^{\kern-#3\ucontract{#1}{#2}\kern #3\kern-#1}}
\def\dcont#1#2#3{_{\kern-#3\dcontract{#1}{#2}\kern #3\kern-#1}}



\font\tenmsy=msbm10
\font\sevenmsy=msbm10 at 7pt
\font\fivemsy=msbm10 at 5pt
\newfam\msyfam 
\textfont\msyfam=\tenmsy
\scriptfont\msyfam=\sevenmsy
\scriptscriptfont\msyfam=\fivemsy
\def\blackB{\fam\msyfam\tenmsy}
\def\Z{{\blackB Z}}

\let\R\rangle

\def\vp{\tilde{\varphi}}
\let\da\dagger

\def\frac#1#2{{\textstyle{#1\over #2}}}

\def\eqalignD#1{
\vcenter{\openup1\jot\halign{
\hfil$\displaystyle{##}$~&
$\displaystyle{##}$\hfil~&
$\displaystyle{##}$\hfil\cr
#1}}
}

\def\eqalignSS#1{
\vcenter{\openup1\jot\halign{
\hfil$\displaystyle{##}$~&
$\displaystyle{##}$\hfil~&
$\displaystyle{##}$\hfil~&
$\displaystyle{##}$\hfil~&
$\displaystyle{##}$\hfil~&
$\displaystyle{##}$\hfil~&
$\displaystyle{##}$\hfil~&
$\displaystyle{##}$\hfil\cr
#1}}
}

\def\text#1{\quad\hbox{#1}\quad}

\def\la{\lambda}

\def\La{\Lambda}

\def\p{\tilde{p}}
\def\P{\tilde{P}}

\def\A{{\cal{A}}}
\def\At{{\tilde{A}}}
\def\Jt{{\tilde{J}}}
\def\Bt{{\tilde{B}}}
\def\B{{\cal{B }}}

\def\y{{\infty}}

\let\Rw\Rightarrow

\def\rw{\rightarrow}

\def\R{\rangle}

\def\su{\widehat{su}}
\def\osp{\widehat{osp}}

\def\p{{\bar p}}

\newcount\eqnum
\eqnum=0
\def\eq{\eqno(\secsym\the\meqno)\global\advance\meqno by1}
\def\eqlabel#1{{\xdef#1{\secsym\the\meqno}}\eq }

\newwrite\refs
\def\startreferences{
 \immediate\openout\refs=references
 \immediate\write\refs{\baselineskip=14pt \parindent=16pt \parskip=2pt}
}
\startreferences

\refno=0
\def\aref#1{\global\advance\refno by1
 \immediate\write\refs{\noexpand\item{\the\refno.}#1\hfil\par}}
\def\ref#1{\aref{#1}\the\refno}
\def\refname#1{\xdef#1{\the\refno}}

\newcount\exno
\exno=0
\def\Ex{\global\advance\exno by1{\noindent\sl Example \the\exno:

\nobreak\par\nobreak}}

\parskip=6pt

\overfullrule=0mm

\def\frac#1#2{{#1 \over #2}}

\let\La=\Lambda

\def\q{{\tilde q}}

\def\uh{{\widehat u}}
\def\rw{{\rightarrow}}

\def\Zt{{\widetilde\Z}}

\newwrite\refs
\def\startreferences{
 \immediate\openout\refs=references
 \immediate\write\refs{\baselineskip=14pt \parindent=16pt \parskip=2pt}
}
\startreferences

\refno=0
\def\aref#1{\global\advance\refno by1
 \immediate\write\refs{\noexpand\item{\the\refno.}#1\hfil\par}}
\def\ref#1{\aref{#1}\the\refno}
\def\refname#1{\xdef#1{\the\refno}}


\Title{\vbox{\baselineskip12pt
\hbox{ }}}
{\vbox {\centerline{ Fermionic characters for graded parafermions}
\bigskip
}}

\centerline{L. B\'egin$^\ddagger$, J.-F. Fortin$^\natural$, P. Jacob$^\natural$ and P.
Mathieu$^\natural$}
\vskip.5cm
\smallskip\centerline{$\ddagger$ \it Secteur des Sciences, Campus
d'Edmundston,}
\centerline{\it Universit\'e de Moncton, Nouveau-Brunswick, Canada, E3V 2S8}
\smallskip\centerline{$^\natural$ \it D\'epartement de Physique,
Universit\'e Laval, Qu\'ebec, Canada G1K 7P4}

\smallskip\centerline{(lbegin@umce.ca,
jffortin@phy.ulaval.ca, pjacob@phy.ulaval.ca, pmathieu@phy.ulaval.ca)}
\vskip .2in
\bigskip
\bigskip
\centerline{\bf Abstract}
\bigskip
\noindent

Fermionic-type character formulae are presented for charged irreducible
modules of the graded parafermionic conformal field
theory associated to the coset
$\osp(1,2)_k/\uh(1)$.
This is obtained by counting the weakly ordered `partitions' subject to 
the graded $\Zt_{k}$ exclusion principle. The bosonic form of the characters is also presented.


\Date{01/03\ }

\let\n\noindent

\newsec{Introduction}

The graded, or $\Zt_k$, parafermionic conformal theory, first introduced in [\ref{J. M. Camino, A. V. Ramallo and
J. M. Sanchez de Santos, Nucl.Phys. {\bf B530} (1998) 715.}\refname\CRS], is equivalent to the coset $\osp(1,2)_k/\uh(1)$
with central charge
$$c=-{3\over 2k+3}\eq$$ In addition to the $\Z_k$ grading of the usual parafermionic models [\ref{A.B. Zamolodchikov and
V.A. Fateev. Sov. Phys. JETP {\bf 82} (1985) 215.}\refname\ZFa], it has an extra $\Z_2$ grading. The basic
parafermionic fields $\psi_{\frac12}$ and $\psi_{\frac12}^\dagger$, which have 
 conformal dimension
$1-1/4k$, are odd with respect to the $\Z_2$ grading. This bi-gradation enforces the relation $(\psi_{\frac12})^{2k}\sim
I$, where $I$ stands for the identity field. 

These theories have been studied further  in [\ref{P.
Jacob and P. Mathieu, Nucl. Phys. {\bf B630} (2002) 433}\refname\JM]. There, two bases of states have been presented. One
is the so-called standard basis, formulated in terms of the modes of $\psi_{\frac12}$ and $\psi_{\frac12}^\dagger$. It is
naturally linked to the graded parafermionic representation theory and, ultimately, to the $\osp(1,2)_k$ representation
theory.  The free action of the modes generates the Verma modules and irreducible modules are obtained by subtracting the
contribution of the singular vectors -- whose explicit expressions have also been presented  in [\JM].

The other basis is a quasi-particle basis. It is formulated  solely from the $\psi_{\frac12}$ modes. Somewhat
surprisingly, the build-in
$\Z_2$ grading prevents a complete ordering of the basis of states. More precisely, the sequence of integers which
corresponds to minus the integer part of the ordered sequence of $\psi_{\frac12}$ modes does not form a partition (which
is defined to be a non-increasing sequence of positive integers). This sequence is only partially ordered: the usual
non-increasing condition applied to adjacent integers $n_i\geq n_{i+1}$ is here replaced by the weaker condition $n_i\geq
n_{i+1}-1$. However, the possible increase by one unit between adjacent integers is limited by the further condition
$n_i\geq n_{i+2}$. In addition, this partially ordered basis is  subject to a constraint that results from  the condition
$(\psi_{\frac12})^{2k}\sim I$. This constraint turns out to be expressible in the form of a generalized Pauli exclusion 
principle. But this is an exclusion of a novel type. As for the usual $\Z_k$ parafermion models, its core
is a difference condition [\ref{P. Jacob and P. Mathieu, Nucl. Phys. {\bf B 620} (2002) 351.}\refname\JMb], but
implemented on mode indices separated by $2k$ instead of $k$.  However the novelty lies in that it is not
formulated as a single condition. It is rather of the form ``A or B must be satisfied''. 

The aim of this paper is to lift these two bases to character expressions. For the standard basis, this leads to
bosonic-type character formulae. Here, the key step amounts to find the generating function for the standard basis.  This
is rather straigthforward because there exists a simple bijection relating the $\Zt_k$ states of the graded standard basis
and those of the usual $\Z_k$ ones.  The remaining step consists in implementing the different singular-vector
subtractions. The result of this construction is presented in section 5. Complementary computations are reported in
appendices A and C (where a detailed example is worked out).

Obtaining the fermionic character is not as simple, however. The difficulty is two-fold: we do not sum over
genuine partitions and the exclusion principle is of a `or'-type. Quite unexpectedly, the method of Andrews [\ref{G.E.
Andrews, Houston J. Math. {\bf 7} (1981) 11.}\refname\Andrr] for summing ordinary partitions with a difference condition
can be adapted to our problem  even though our recurrence relations are much more complicated.  Mimicking his argument led
us almost directly to the sought for generating function. This analysis is presented in section 3. Using then a standard
procedure (cf. [\ref{J. Lepowsky and M. Primc,  Contemporary Mathematics {\bf 46} AMS, Providence, 1985.}\refname\LP]),
this function can be transformed into character expressions, presented in section 4. Section 2 is devoted to reviewing the
results of [\JM] pertaining to the quasi-particle basis.

\newsec{Reviewing the $\Zt_k$ quasi-particle basis}

The quasi-particle basis of the graded parafermions  is formulated
in terms of strings of operators $\B$, where
$$\B_{-n} |\Phi_\q\R\equiv  B_{-n+(1+2\q)/4k}|\Phi_\q\R\eqlabel\didi$$
where $|\Phi_\q\R$ stands for a generic state of charge $\q$ and  the operator $B$ is the mode of the basic parafermion
$\psi_{\frac12}$.   The conformal dimension of a $\B$ mode is
read off from the negative of the full subindex of the corresponding $B$ mode: on a state of charge $q$ it is
$n-(1+2\q)/4k$ (cf. eq. (\didi)). Note that
$\B$ itself has charge 1, so that, e.g.,
$$\B_{-n_1}\B_{-n_2} |\Phi_\q\R\equiv  B_{-n_1+(1+2\q+2)/4k} B_{-n_2+(1+2\q)/4k}|\Phi_\q\R\eq$$
i.e., the omitted fractional part of a $\B$ term within a string is position dependent.
Let $|\vp_\q\R$ stands for a graded parafermionic highest weight; it satisfies:
$$\B_{p} |\vp_\q\R=0 \qquad p\geq 0\eqlabel\hwcon$$
and its conformal dimension reads:
$$h_{\vp_\q}= {\q(2k-3\q)\over 4k (2k+3)}\eqlabel\pridi$$
We are interested in the completely reducible representations, which correspond to the case where $\q$ is an integer
in the range
$0\leq \q\leq k$.

The quasi-particle basis is described in terms of {\it partially ordered}
$\B$-strings  of the form
$$\B_{-n_1}\B_{-n_2}\cdots \B_{-n_m}|\vp_\q\R\eqlabel\strib$$ 
 constrained by the 
{\it weak-ordering
conditions}:
$$ n_i\geq n_{i+1}-1,\qquad n_i\geq n_{i+2} ,\qquad n_i\geq 0
\eqlabel\oror$$
The vector $(n_1,\cdots, n_m)$ does not form a  genuine partition because it does not satisfy
the non-decreasing condition.  Indeed, an increment by one unit from $n_i$ to $n_{i+1}$ is allowed.
However a further increase by one unit from $n_{i+1} $ to $n_{i+2}$ is ruled out. Due to the jagged (or serrated)
nature of the lowest-weight `partition' $(\cdots 01010101)$, a vector
$(n_1,\cdots, n_m)$ satisfying (\oror) will be called a {\it jagged partition}.

For the characterization of the
states in a highest-weight module, 
 we also need to take into account the singular vector $
(\B_0\B_{-1})^{k-\q+1}|\vp_\q\R$, which is captured by the boundary condition
$$n_{m-2k+2\q}\geq 2\qquad {\rm or } \qquad   n_{m-2k+2\q-1}\geq 1\eqlabel\bdry$$ 
Indeed,
a sequence of pairs $\B_0\B_{-1}\cdots\B_0\B_{-1}$ is either ended by
$\B_{-2}$ or $\B_{-1}\B_{-1}$ acting on the left.

As a result of the model's $\Zt_{k}$ invariance, $\B$ strings are further
subject to a generalized form of exclusion principle [\JM], which  translates into the following difference
conditions:
$$\eqalignD{
& \quad &n_i \geq  n_{i+2k-1} +1\cr & & \qquad{\rm or} \cr 
&~& n_i = n_{i+2k-1} \quad {\rm and}
\quad n_{i+1} =  n_{i+2k-2}+2 \cr}\eqlabel\allo$$ 
for all values of $i\leq m-2k+1$.  

The goal is thus to count the number of jagged partitions subject to the
constraints (\bdry) and (\allo) (called $\Zt_{k}$ jagged partitions for short).
Actually, we are interested in solving  a more precise problem, which is to construct the
character of  modules with specified charge.  Recall that the charge is defined modulo
$2k$ so that the module of highest-weight
$|\vp_\q\R$ of relative charge $r$ is obtained by summing up all allowed
$\B$ strings of length $r$ modulo $2k$. Our main objective is thus to  sum over all $\Zt_{k}$ jagged partitions of
length
$m=2kj+r$ where $j$ is a non-negative integer.  Without lost of generality, $r$ will also be taken to be non-negative.

The top state in the  module $|\vp_\q\R$ of relative charge $r$
is 
$$|\vp_\q^{(r)} \R\equiv \B_{-i}\cdots \B_{0}\B_{-1}\B_{0} \B_{-1}|\vp_\q\R\eq$$
with $i=0$ (1) if $r$ is even (odd). The dimension of this state is
$$h_{\vp_\q^{(r)}}= h_{\vp_\q}+ \left[{r+1\over 2}\right]-{r(r+2\q)\over 4k}\eq$$
where $[x]$ stands for the largest integer smaller than $x$. On the other hand, taking into
account the hidden fractional part of the mode indices, the conformal dimension of the $\B$-string (\strib) is
found to be
$$h_{\vp_\q}+ \sum_{i=1}^mn_i -{m(m+2\q)\over 4k}\eq$$
If we redefine this dimension with respect to the state $|\vp_\q^{(r)}\R$, and set $m=2kj+r$, we
have
$$h_{\vp_\q^{(r)}} + \sum_{i=1}^mn_i -j(kj+r+\q)-\left[{r+1\over 2}\right]\eqlabel\reca$$
In other words, the grade $s$ of the state (\strib) (with  $m=2kj+r$) in the  $|\vp_\q^{(r)}\R$ module is
$$s= \sum_{i=1}^mn_i -j(kj+r+\q)-\left[{r+1\over 2}\right]\eqlabel\recass$$


\newsec{Counting $\Zt_{k}$ constrained jagged partitions}

\subsec{Partition recurrence relations}

Let us introduce the following  sets of $\Zt_{k}$ jagged partitions with prescribed boundary conditions:

\n $A_{k,2i}(m,n)$:  $\#$ of $\Zt_{k}$ jagged partitions of $n$ into $m$  parts with {\it at most} $(i-1)$
pairs of ``01" at the right (with $1\leq i\leq k$).

\n $ B_{k,j} (m,n)$: $\#$ of $\Zt_{k}$ jagged  partitions of $n$  into $m$  parts with {\it at
most}Ê $(j-1)$Ê``1" at the right (with $1\leq j\leq 2k$) .\foot{It is  understood that these sets are exclusive, i.e.,
that $B_{k,j} (m,n)$ excludes all partitions containing 0, despite  terminating with
a ``1''.}
 
These sets satisfy the following recurrence relations: 
$$\eqalignD{ 
&(i)\qquad  & A_{k,2i} (m,n)-A_{k,2i-2 }(m,n)= B_{k,2k-2i+2} (m-2i+2,n-i+1)\cr
&(ii) \qquad& B_{k, 2i+1 } (m,n) - B_{k, 2i } (m,n)=Ê A_{k, 2k-2i} (m-2i,n-m) \cr
&(iii)\qquad & B_{k, 2i } (m,n) - B_{k, 2i-1 } (m,n)=Ê A_{k, 2k-2i+2} (m-2i+1,n-m) \cr
}\eq$$
together with the boundary conditions:
$$\eqalignD{
&(iv)\qquad  & A_{k,2}(m,n)=B_{k,2k}(m,n)\cr
&(v)\qquad  & A_{k,2k}(m,n-m)=B_{k,1}(m,n)\cr
&(vi)\qquad & A_{k,2i}(0,0)= B_{k,i}(0,0) = 1\cr}  \eq$$
Moreover, both $A_{k,2i}(m,n)$ and $B_{k,i}(m,n)$ are assumed to be 0 if any of the argument $n$ or $m$ is
negative or if one of them is 0. In addition we impose
$$ A_{k,0}(m,n)=B_{k,0}(m,n)=0\eq$$

These recurrence relations are obtained as follows\foot{These derivations are much inspired by the
analogous ones of Andrews in [\Andrr] (see also
[\ref{G.E. Andrews, {\it The theory of
partitions}, Cambridge Univ. Press (1984).}\refname\Andr]),  
pertaining to ordinary partitions $(\la_1,\cdots,  \la_m) $
subject to the difference condition $\la_i\geq \la_{i+k-1}+2$.}.  For all three cases, the left hand side is
designated to isolate sets of jagged  partitions with a precise boundary term. In particular, $A_{k,2i}
(m,n)-A_{k,2i-2 }(m,n) $ yields the number of
$\Zt_{k}$ jagged  partitions of $n$ into $m$  parts containing {\it  exactly} $i-1$
pairs of ``01" at the right. Taking out the tail 
 $01\cdots 01$, reducing then the length of the partition from $m$ to $m-2(i-1)$
and its weight $n$ by $i-1$, we end up with  $\Zt_{k}$ jagged partitions which can terminate
with a certain number of ``1''. But this number is constrained by the exclusion principle.
Before taking out the tail, the number of successive ``1''  is at most $2k-2(i-1)-1$, which gives the right hand side of
$(i)$. Note that this is the very place where the exclusion does enter in this recurrence relation.  The other two
relations
$(ii)-(iii)$ are proved in the same way. Note that $(i)$ codes the exclusion of the basic strings
$((11)^{k-j}(01)^j)= (11\cdots1101\cdots 01)$ while the other two relations code the exclusion of the set
$((12)^{k-j}(11)^j)$ (where $1\leq j\leq k$ in both cases).\foot{These are the elementary excluded strings; the
corresponding generic form of the excluded
$\B$ strings are
$(\B_{-n}\B_{-n})^{k-j} (\B_{-{n}+1}\B_{-{n}})^j$ and $(\B_{-n}\B_{{-n-1}})^{k-j} (\B_{-{n}}\B_{-n})^j$
(which corrects an error in the first equation of (5.3) in [\JM]).}

Consider next the boundary conditions $(iv)-(vi)$. Obviously, if there are no ``01'', there can be at most $2k-1$ ``1''
at the end of the string, which gives $(iv)$. On the other hand, if there are no ``01'' nor ``1'' at the end,
we can strip off the string by the basic $m$-string $11\cdots 11$: this reduces the value of $n$ by $m$ and the
resulting string can have any number of ``01'' allowed by the exclusion: this yields $(v)$.\foot{It should be clear
that the boundary conditions $(iv)$ and $(v)$ are in fact consequences of $(i)$ and $(ii)$ respectively for $i=1,0$.}

As an illustrative example, consider strings of length 6 and weight 8 for $k=3$.
The set $A_{3,4}(6,8)$ (with at most one pair of ``01'' at the end) is: 
 $$\eqalign{ 
  &(4, 1, 1, 1, 0, 1)\quad (3, 2, 1, 1, 0, 1) \quad (3, 1, 2, 1, 0, 1) \quad (3, 1, 1, 1, 1, 1)
 \quad(2, 3, 1, 1, 0, 1)\cr &(2, 2, 2, 1, 0, 1) \quad(2, 2, 1, 2, 0, 1) \quad(2, 2, 1, 1, 1, 1)
 \quad( 2, 1, 2, 1, 1, 1)\cr}\eq$$
Similarly, the full set of elements of $A_{3,2}(6,8)$ is
$$
(3, 1, 1, 1, 1, 1)\quad
(2, 2, 1, 1, 1, 1)\quad
(2, 1, 2, 1, 1, 1)\eq$$
The elements of $A_{3,4}(6,8) - A_{3,2}(6,8)$, with their ``01'' tail cut off, are
$$(4, 1, 1, 1)\quad 
(3, 2, 1, 1)\quad
(3, 1, 2, 1)\quad
(2, 3, 1, 1)\quad
(2, 2, 2, 1)\quad
(2, 2, 1, 2)\eq$$
and these are indeed all the elements of 
$B_{3,4}(4,7)$.

\subsec{Generating functions}

Let us define the
generating functions:
$$\eqalign{
& \At_{k,2i}(z;q)= \sum_{m,n\geq 0} z^mq^n A_{k,2i}(m,n)\cr
& \Bt_{k,i}(z;q)= \sum_{m,n\geq 0} z^mq^n B_{k,i}(m,n)\cr}\eq$$
The recurrence relations are now lifted to the following functional relations:
$$\eqalignD{ 
&(i)'\qquad  & \At_{k,2i} (z;q)-\At_{k,2i-2 }(z;q)= z^{2i-2}q^{i-1}\Bt_{k,2k-2i+2}(z;q)\cr 
&(ii)'
\qquad & \Bt_{k, 2i+1 } (z;q) - \Bt_{k, 2i } (z;q)=Êz^{2i}q^{2i} \At_{k, 2k-2i} (zq;q)
\cr
&(iii)'\qquad & \Bt_{k, 2i } (z;q) - \Bt_{k, 2i-1 } (z;q)=Êz^{2i-1}q^{2i-1} \At_{k, 2k-2i+2} (zq;q)
\cr }\eq$$ while the boundary conditions become:
$$\eqalignD{
&(iv)'\qquad  & \At_{k,2}(z;q)=\Bt_{k,2k}(z;q)\cr
&(v)'\qquad  & \At_{k,2k}(zq;q)=\Bt_{k,1}(z;q)\cr
&(vi)'\qquad & \At_{k,2i}(0;q)= \At_{k,2i}(z;0)=\Bt_{k,i}(0;q) =\Bt_{k,i}(z;0) =1\cr}  \eq$$

The solutions of these recurrence relations are
$$\eqalign{
& \At_{k,2i}(z;q)= (-zq)_\y\, F_{k,i}(z^2;q)\cr
& \Bt_{k,2i+1}(z;q)= (-zq)_\y \,F_{k,i}(z^2q;q)+z^{2i}q^{2i}(-zq^2)_\y \,F_{k,k-i}(z^2q^2;q)\cr
& \Bt_{k,2i}(z;q)= (-zq)_\y \,F_{k,i}(z^2q;q)\cr
}\eqlabel\solrecu$$
where
$$(a)_n=(a;q)_n= \prod_{i=0}^{n-1} (1-aq^i)\eqlabel\dean$$
Indeed, the relation $(i)'$ becomes
$$F_{k,i}(z^2;q)- F_{k,i-1}(z^2;q) = z^{2i-2}q^{i-1} F_{k,k-i+1}(z^2q;q)\eq$$
whose solution is known to be [\Andrr] (cf. the proof of theorem 1, with $d=0$ and $z$ replaced by $z^2$):
$$F_{k,i}(z^2;q)= \sum_{m_1,\cdots,m_{k-1}=0}^\y {
q^{N_1^2+\cdots+ N_{k-1}^2+L_{i}}\; z^{2N} \over (q)_{m_1}\cdots (q)_{m_{k-1}}
}\eqlabel\ande$$ where $$\eqalignD { & N= N_1+\cdots+N_{k-1}\, , \qquad &N_i= m_i+\cdots
+m_{k-1}\cr & L_j=N_j+\cdots N_{k-1}\,, & L_k=L_{k+1}=0 \cr}\eq$$
More precisely, this is the only solution  satisfying $F_{k,i}(z^2;0)= F_{k,i}(0;q)= 1$ and
$F_{k,1}(z^2;q)=F_{k,k}(z^2q;q)$. This takes into account the conditions $(iv)'$ and part of $(vi)'$. The other
relations do not impose further constraints.

However, the above verification is only partial in that it leaves a possible undetermined prefactor (that must reduce
to 1 when either $z$ or $q$ vanishes) common to both
$\At(z;q)$ and
$\Bt(z;q)$. 
But this  prefactor has to be $k$ independent. The
$k$-independence is rooted in  the following relation between $F_{k,i}$ and $F_{k-1,i}$ [\Andrr] (cf. eq. (2.11)):
$$F_{k,i}(w;q)= \sum_{n\geq 0}{q^{(k-1)n^2+(k-i)n}\, w^{(k-1)n}\over (q)_n}\, F_{k-1,i}(wq^{2n};q)\eq$$
Since this expression satisfies the $F_{k,i}(w;q)$ recurrence relation, to maintain the validity of these
relations for the product of $F_{k,i}(w;q)$ with a prefactor, it is clear that this
prefactor has to be
$k$ independent.  It can thus be fixed by setting $k=2$.

Let us then solve the recurrence relations explicitly for $k=2$. The various relations are:
$$\eqalign{
& \At_{2,4}(z;q)= \At_{2,2}(z;q)+z^2q\Bt_{2,2}(z;q)\cr
& \Bt_{2,4}(z;q)= \Bt_{2,3}(z;q)+z^3q^3\At_{2,2}(zq;q)\cr
& \Bt_{2,3}(z;q)= \Bt_{2,2}(z;q)+z^2q^2\At_{2,2}(zq;q)\cr
& \Bt_{2,2}(z;q)= \Bt_{2,1}(z;q)+zq\At_{2,4}(zq;q)\cr
& \Bt_{2,4}(z;q)= \At_{2,2}(z;q)\cr
& \Bt_{2,1}(z;q)= \At_{2,4}(zq;q)\cr}\eq$$
By eliminating the terms $\Bt_{2,j}$ and $\At_{2,2}$, we end up with
$$\At_{2,4}(z;q)= (1+zq)(1+z^2q+z^2q^2)\At_{2,4}(zq;q) -z^4q^5 (1+zq)(1+zq^2)\At_{2,4}(zq^2;q)\eqlabel\aqaq$$
Setting 
$$\At_{2,4}(z;q)= (-zq)_\y C_{2,4}(z;q)\eqlabel\acce$$
(where $(-zq)_\y$ is defined as the limit $n\rw\y$ of (\dean))
the infinite product cancels on both sides and we get 
$$C_{2,4}(z;q)= (1+z^2q+z^2q^2)C_{2,4}(zq;q) -z^4q^5 C_{2,4}(zq^2;q)\eq$$
Looking for a  solution of the form
$$C_{2,4}(z;q)= \sum_{n=0}^\y a_n(q) z^n\eq$$ leads to the recurrence relation:\foot{The dramatic simplification that
results from the  factorization the prefactor $(-zq)_\y$ should be stressed; without this factorization, we would
end up with a six-term recurrence relation.}
$$a_n={q^{n-1}(1+q)\over 1-q^n}\, a_{n-2}-{q^{2n-3}\over 1-q^n}\, a_{n-4}\eq$$with $a_0=1$  and $a_{i<0}=0$.  The
later relation immediately forces the vanishing of all the coefficients  $a_{2n+1}$. Those with even indices
are found to be
$$a_{2n}={q^{n^2}\over (q)_n}\qquad \Rw\qquad C_{2,4}(z;q)= \sum_{n=0}^\y {q^{n^2}\over (q)_n} z^{2n}\eqlabel\accec$$
Therefore,  $C_{2,4}(z;q)=F_{2,2}(z^2;q)$ and this proves that the prefactor relating $\At_{k,2i}(z;q)$ to
$F_{k,i}(z;q)$ is indeed $(-zq)_\y$.  This completes the proof of (\solrecu).

It is quite remarkable that the transition from counting $\Z_k$ constrained partitions to counting $\Zt_k$ constrained
jagged partitions is fully captured by the replacement of $z \rw z^2$ in the $\Z_k$
generating function and the introduction  of the prefactor $(-zq)_\y$. The heuristic rationale for the transformation
$z \rw z^2$ (recalling that the power of $z$ keeps track of the length) is that $\Zt_k$ jagged partitions refer to
$\B$-strings while $\Z_k$ partitions refer to $\A$-strings and that $\B\B\sim \A$.

Using $$(-zq)_\y= \sum_{m=0}^\y {z^mq^{m(m+1)/2}\over (q)_m}\eq$$
we can rewrite the resulting expression of $\At_{k,2i}(z;q)$  in a compact form
$$\At_{k,2i}(z;q)=  \sum_{m_0,\cdots,m_{k-1}=0}^\y {
q^{m_0(m_0+1)/2+N_1^2+\cdots+ N_{k-1}^2+L_{i}}\; z^{m_0+2N} \over (q)_{m_0}\cdots (q)_{m_{k-1}}
}\eqlabel\andee$$

With $z$ set equal to 1, the above sum can be reexpressed in product form as:
$$
\eqalign{
\At_{k,2i}(1;q) &= (-q)_\y\prod_{n=1\atop n\not=0,\pm i ~{\rm mod}~(2k+1)}^\y{1\over (1-q^n)}\cr
&= {1\over (q;q^2)_\y} \prod_{n=1\atop n\not=0,\pm i ~{\rm mod}~(2k+1)}^\y {1\over (1-q^n)}\cr}\eq$$
The first step uses the Andrews-Gordon identity [\Andr] (cf. theorem 7.8). Again, up to the prefactor, this is
precisely the product form of the usual
$\Z_k$ models.

\newsec{Fermionic-type $\Zt_{k}$ characters}

Let us now adapt the counting of $\Zt_k$ jagged partitions to our problem of 
constructing the character of relative
charge
$r$ over the module
$|\vp_\q\R$.  At first, the singular-vector constraint fixes  $i-1=k-\q$.
Second, constructing a fixed-charge character amounts to  sum only over jagged partitions of
lengths equal to
$r$ mod
$2k$. To achieve this restriction, one could simply sum over all values of $m$ but set
$z^m=0$ unless
$m=2kj+r$  (see [\LP] section 6 and also [\JMb] section 5). When $m=2kj+r$,
we choose $z$ to be related to $q$ in such a way that the total power of $q$ is reduced from $n$ to the level $s$
counted from the state $|\vp_\q^{(r)}\R$, that is,
$s=n-j(kj+r+\q)-[(r+1)/2]$ -- cf. (\reca). We thus need
 $$ z=f(q)\quad{\rm with}\quad [f(q)]^m\equiv\left\{ \matrix{ q^{-j(kj+r+\q)-[(r+1)/2]}\quad& {\rm
when}\quad m=2kj +r \cr 0 &{\rm otherwise}\cr} \right.\eqlabel\zmdef$$

The $\Zt_{k}$ character over the module $|\vp_\q\R$ with relative charge $r$ is thus
$$\chi_{\vp_\q^{(r)}}(q)= q^{h_{\vp_\q^{(r)}}-c/24} \sum_{n,j=0}^\y q^s
A_{k,2(k-\q+1)}\big(2kj+r,s+j(kj+r+\q)+[(r+1)/2]\big) \eq$$ or equivalently
$$\chi_{\vp_\q^{(r)}}(q)= q^{h_{\vp_\q}^{(r)}-c/24} \At_{k,2(k-\q+1)}(f(q);q)
\eq$$
with $f(q)$ given in (\zmdef). Up to the prefactor $q^{h_{\vp_\q}^{(r)}-c/24} $, the character is
thus
$$ \sum_{m_0,\cdots,m_{k-1}=0\atop m_0+2N=\,r~{\rm mod} ~2k}^\y {
q^{\frac12m_0(m_0+1)+N_1^2+\cdots+ N_{k-1}^2+L_{k-\q+1}-\frac1{4k}(m_0+2N-r)(m_0+2N+r+2\q)-[\frac12(r+1)]}
\over (q)_{m_0}\cdots (q)_{m_{k-1}} }\eqlabel\andeee$$
This is our main result.

Note that for $k=1$, which corresponds to the minimal model ${\cal M}(3,5)$ [\JM], this expression is substantially
simplified. The various characters read then:
$$\eqalign{
&\chi_{\vp_0^{(0)}}(q)=\chi_{1,1}^{(3,5)}(q) = q^{1/40} \sum_{n\geq 0}{q^{n(n+1)}\over (q)_{2n}}\;,\cr 
&\chi_{\vp_0^{(1)}}(q)=\chi_{{1,4}}^{(3,5)}(q) = q^{31/40} \sum_{n\geq 0}{q^{n(n+2)}\over (q)_{2n+1}}\;,\cr
&\chi_{\vp_1^{(0)}}(q)=\chi_{{1,2}}^{(3,5)}(q) = q^{-1/40} \sum_{n\geq 0}{q^{n^2}\over (q)_{2n}}\;,\cr 
&\chi_{\vp_1^{(1)}}(q)=\chi_{{1,3}}^{(3,5)}(q) = q^{9/40} \sum_{n\geq 0}{q^{n(n+1)}\over (q)_{2n+1}}\cr}\eq$$
which reproduce the already known fermionic formulae of [\ref{R. Kedem, T.R.
Klassen, B. M. McCoy and E. Melzer, Phys. Lett. {\bf B307} (1993) 68.}\refname\KKMMa].

\newsec{Bosonic $\Zt_k$ characters}

\subsec{The $\Zt_k$ standard basis and the Verma module character}

In view of constructing the bosonic characters, whose corresponding 
irreducible modules are given  by alternating
sums of Verma modules, we need to find the character of a highest-weight Verma module. 
A Verma module is obtained
from the free action of all the modes of the complete algebra generators acting on a 
highest-weight.
In the present context, the complete algebra generators refers to the basis parafermions
$\psi_{\frac12}$ and $\psi_{\frac12}^\dagger$ (whose modes are denoted as $\B^\dagger$). The basis
of states describing the various states obtained from the free action of the mode $\B$ and
$\B^\dagger$ is called the standard
basis.

The standard basis can be described as follows [\JM]:
$$ \B_{-n_1}\B_{-n_2}\cdots \B_{-n_p}\B^\da_{-n'_1}\B^\da_{-n'_2}\cdots \B^\da_{-n'_{p'}}|\vp_\q\R\eqlabel\stada$$ 
with the weak ordering condition (\oror) applying separately to the $n$'s and to the $n'$'s, with 
$n_p\geq 1, \;
n'_{p'}\geq 0$ and $p-p'=r$ (recall that $\B^\da$ has charge $-1$). We stress that in the standard basis, the charge
assignment is  absolute, i.e., it is not defined modulo $2k$. The condition
$n'_{p'}\geq 0$ comes from the highest-weight condition 
$$\B^\da_{m} |\vp_\q\R  =0\qquad m\geq 1\eqlabel\hwdag$$
which, in addition to (\hwcon ), fully characterize a highest-weight state [\CRS,\JM].

How do we count states in the standard basis? Consider first the case where there are only $\B$ strings. The
contributions of these strings to the character of relative charge $r$  is obtained by counting jagged partitions of
length {\it exactly equal to }
$r$. Similarly, it would seem that the contributions of $\B^\dagger$ strings to the  character of relative charge $-r$ 
are obtained by counting jagged partitions of length
$r$. But this is not quite so. Indeed, since the highest-weight condition (\hwdag) allows for $\B^\dagger$ zero modes,
we actually need to sum over all jagged partitions of length {\it at most} 
$r$.\foot{Let us make a little clarification concerning the length of jagged partitions. For an ordinary
partition, the length is the number of non-zero entries. Jagged partitions however do contain essential zero entries.
The length is obtained by dropping the right-most tail of zeros; it is thus the number of entries counted from the
right-most positive number. For instance,
$323120101000$ has length 9. But this jagged partition (with its tail of zeros) would contribute to the set of those
describing strings containing 12 
$\B^\da$ operators.} 

Let us denote by $J(m,n)$ the set of jagged partitions of length $m$ and weight $n$ (with
$J(0,0)=1$ and $J(0,n>0)=0$). Similarly, we denote by $J(\leq m,n)$ the set of jagged partitions
of length at most
$m$. Manifestly, 
$$J(\leq m,n)=\sum_{m'=0}^m J(m',n)\eq$$
Note that for jagged partitions there is no one-to-one correspondence (analogous to the one that exists for ordinary
partitions) between the sets $J(\leq m,n)$ and $J(m,n')$ where $n'$ is an appropriately shifted value of the weight
$n$.\foot{ Recall that an ordinary partition of weight $n$ and length at most  $m$ can be transformed into a partition of
length exactly equal to $m$ by adding  $m$ ``1''; this increases the weight by
$m$. Take for instance
$m=4$: (41) is mapped into (5211) by adding (1111). However, there is no such relation for jagged partitions. In that
case, it would have been natural to add the fundamental partition $(i\cdots 010101)$ where $i=[1-(-1)^m]/2$ to a
jagged partition of length  smaller than $m$ to make its length precisely equal to $m$ (obtaining thereby an element
of $J(m,n+[(m+1)/2])$). However, a simple example readily shows the incorrectness of this procedure. Take the set
$J(\leq 3,3)= \{(3), \, (21), \, (12),\,(111),\,(201)\}$; adding to each element the string (101) yields
$\{(401),\,(311),\,(221),\,(212),\,(302)\}$.
Manifestly, (302) is not a jagged partition (the condition $n_2\geq n_3-1$ being violated). Indeed, the set $J(3,5)$
contains only the first four elements of the above list.  This counter-example reflects the plain fact that the sum of
two jagged partitions is not itself a jagged partition. }

For states containing both $\B$ and $\B^\dagger$ strings, since the $\B$ and $\B^\da$ substrings are ordered
separately, the only matching requirement comes from the fixed value of the total relative charge, which correlates the
lengths of the two strings. The counting of states at level $s$ in the module of relative charge $r$ is thus\foot{This
holds for $r\geq 0$ as well as for $r<0$. In the later case, we note that $J(m-|r|,n)=0$ unless $m\geq |r|$.}
$$G_r(s)=\sum_{s_1+s_2=s\geq 0}\, \sum_{m\geq 0}\,  \sum_{m'=0}^m J(m+r,s_1)J(m',s_2)
\eqlabel\grr$$
 The character of the Verma module over $|\vp_\q\R$ with charge $r$ is thus
$$\chi_{\vp_\q^{(r)}}(q)= q^{h_{\vp_\q^{(r)}} -c/24}\, q^{-[(r+1)/2]}\, V_{r}(q)\eq$$ with
$$V_r(q)\equiv \sum_{s=0}^\y q^s G_{r}(s)\eq$$
i.e., the exponent ${h_{\vp_\q^{(r)}}}$ of the prefactor captures the whole dependence upon the choice of the
highest-weight state, in particular, its charge.  The extra factor $q^{-[(r+1)/2]}$ is forced by the fact that the leading
power of $V_r$ $(r\geq 0)$ is precisely  $q^{[(r+1)/2]}$.

\subsec{Irreducible character: subtracting the singular vectors}

In [\JM], we  found the explicit expression of all the singular vectors. These are
$$\eqalign{
 |\La_{\q,\ell}^{\rm I(i)}\R &=  \; (\B_0^\da)^{2\ell\la+\mu}(\A_{-1})^{\frac12((2\ell-1)\la+\mu)} \cdots
(\A_{-1})^{\frac12(\la+\mu)} 
(\B_0^\da)^{\mu} |\vp_\q\R\cr
|\La_{\q,\ell}^{\rm I(ii)}\R&=  \; (\A_{-1})^{\frac12((2\ell+1)\la+\mu)} |\La_{\q,\ell}^{\rm I(i)}\R\cr
|\La_{\q,\ell}^{\rm II(i)}\R &= \; (\A_{-1})^{\frac12((2\ell+1)\la-\mu)}\cdots
 (\B_0^\da)^{2\la-\mu} (\A_{-1})^{\frac12(\la-\mu)}|\vp_\q\R\cr
|\La_{\q,\ell}^{\rm II(ii)}\R 
&=\; (\B_0^\da)^{(2\ell+2)\la-\mu} |\La_{\q,\ell}^{\rm II(i)}\R \cr
}\eqlabel\sisi$$
where $$ \la= 2k+3\; ,\qquad \mu = 2\q+1\eq$$
(Note that the power of $\A_{-1}$ is always integer.)
For both types of  singular vectors, we have distinguished the cases (i)
where there is an odd number of groups of terms (a group refers to a power of
$\A_{-1}$ or
$\B^\da_0$ and will often be called a {\it factor} for short), from cases
(ii) which have an even number of factors.

It is easily checked that for all groups of terms of the form ${\cal C}^{d+1}$ appearing in the expression of the
singular vectors, the  sum of the conformal dimension of the rightmost $d$ terms add up to zero.  The conformal
dimension of the group is thus given by that of the leftmost term and it is simply $-(d+1)/(\epsilon k)$ with
$\epsilon=1$ if ${\cal C}=\A$ and $4$ if ${\cal C}=\B^\da$. We can thus read off the dimension of the singular vector
$|\La\R$ simply from the total sum of the various powers of
$\A_{-1}$ or
$\B^\da_0$ factors that it contains:
$$h_{\La}= h_{\vp_\q}-\left({\# \A_{-1}\over k} +{\# \B^\da_0\over 4k}\right)\eqlabel\dirm$$
The number of modes in each singular vector is
$$\eqalignD{
 & |\La_{\q,\ell}^{{\rm I(i)}}\R:\qquad \# \A_{-1} =a_{\ell-1} &\qquad \# \B^\da_{0}= b_\ell \cr
&|\La_{\q,\ell}^{{\rm I(ii)}}\R:\qquad \# \A_{-1}= a_{\ell} &\qquad \# \B^\da_{0}= b_\ell\cr
&|\La_{\q,\ell}^{{\rm II(i)}}\R:\qquad \# \A_{-1}= a'_{\ell} &\qquad \# \B^\da_0= b'_{\ell-1}\cr
&|\La_{\q,\ell}^{{\rm II(ii)}}\R:\qquad \# \A_{-1}= a'_{\ell}&\qquad\# \B^\da_0= b'_\ell \cr}
\eqlabel\sytb$$
with
$$\eqalignD{
& a_\ell= \frac12(\ell+1)[(\ell+1)\la+\mu]\qquad  &b_\ell= (\ell+1)[\ell\la+\mu] \cr
& a'_\ell= \frac12(\ell+1)[(\ell+1)\la-\mu] \qquad  & b'_\ell=
(\ell+1)[(\ell+2)\la-\mu]\cr}\eqlabel\abab$$ (Note that $a'_\ell$ is obtained from $a_\ell$ by reversing the sign of
$\mu$ but this relation does not hold for the
$b$ expressions.)

At first sight, it could seem paradoxical that the dimension of these singular vectors be always {\it smaller} than
that of the original highest-weight state. The point, however, is that in building an irreducible character of a given
charge, we need to take into account those descendents of these singular vectors which have the proper charge. In
other words, we have to `fold' the singular vectors within the module of interest, by acting on them with the
appropriate number (modulo $2k$) of $\B$ or $\B^\da$ operators.  These folded descendents, which would need to be
subtracted, have dimension larger than that of the  highest-weight state. 

In view of computing the dimension of the singular descendants in the simplest way, let us evaluate the fractional
dimension of a generic state of the form (\stada). It is not difficult to see that if $p=p'$, the total fractional
dimension of the $\B$ operators cancels exactly that of the $\B^\da$ ones. On the other hand, if $p-p'=r$, there is a
residual fractional dimension that needs to be taken into account.  But this is precisely taken into account when
evaluating the value of the level with respect to the charged highest-weight state $|\vp_\q^{(r)}\R$, e.g., this
residual value is equal to $h_{\vp_\q^{(r)}}- h_{\vp_\q}$.  The net effect is that we do not need to care anymore about
the fractional parts to calculate the dimension of the folded singular vectors. The first subtracted state is thus at
level given by the total number of
$\A_{-1}$ operators in the singular vectors (which has been computed above), added to the number of $\A_{-1}$ operators
required in the folding step.\foot{For this lowest dimensional folded state, the number of $\B^\da$ operators does not
enter in the computation of the dimension since they have zero integer mode.} 

To build up the character formula for $\vp_\q$, we need to subtract the singular
vectors of type (i) and add those of type (ii), once folded in the sector of charge $r$ (which,  as already said, is
regarded to be positive).  Denote by
$V_{r}$ the free module of relative charge $r$.  The irreducible character is thus
$$\chi_{\vp_\q^{(r)}}(q)= q^{h_{\vp_\q^{(r)}} -c/24}q^{-[(r+1)/2]} M_{\q,r}(q)\eqlabel\cabo$$
(again, the overall prefactor
$q^{-[(r+1)/2]}$ is inserted to ensure that the leading term of $M_{\q,r} $ is of order $q^0$), where
$$\eqalign{
M_{\q,r}(q)  = \{ V_{r} + \sum_{\ell=0}^\y  [&-q^{a_{\ell-1}} V_{\ell\la+\mu+r}+q^{a_\ell}
V_{-(\ell+1)\la+r}\cr & -q^{a'_\ell}V_{-(\ell+1)\la+\mu+r}+q^{a'_\ell}ÊV_{(\ell+1)\la+r}]\}\cr} 
\eqlabel\caboo$$
Note that the $\q$-dependence of $M_{\q,r}$ is fully captured in the $\mu$ factors ($\mu=2\q+1$).  The power of
$q$ in front of each term is indeed the number of
$\A_{-1}$ factors contained in each singular vector (whose contributions appear in the order I(i), I(ii), II(i), II(ii)).
The charge of each Verma module appearing in the sum is given by the number of $\B_0^\da$ minus twice the number of 
$\A_{-1}$ operators of the corresponding singular vectors, plus $r$. For instance, 
$|\La_{\q,\ell}^{\rm I(i)}\R$ has charge $2a_{\ell-1}-b_\ell= -\ell\la-\mu$. Therefore, the descendents of
$|\La_{\q,\ell}^{\rm I(i)}\R$ that  contribute to the charge-$r$ module
are those of charge $\ell\la+\mu+r$. This explains the subindex of the first
term within the sum. The other ones are obtained in the same way.


\subsec{Counting unconstrained jagged partitions }

In order to get an explicit form for the Verma-module character, we now 
count directly the number of states in a Verma module, i.e., find their generating function. This amounts to sum over
jagged partitions
 without imposing any
exclusion constraints, nor special boundary conditions. This can be achieved in various ways. 
The simplest method consists in 
taking directly the limit $k\rw \y$ of the generating function $\At_{k,2k}(z;q)$ (the choice of $i=k$ readily takes
away any boundary constraint). This is the approach used in this section. Two different ones are presented in appendix A.

The function $G_r(s)$ defined in (\grr) is a bilinear sum of $J(m,n)$'s. It can be constructed by appropriately fusing
two generating functions for these $J(m,n)$ as explained below. The first step is thus to compute this generating
function, defined as
$$\Jt(z;q)= \sum_{n,m=0}^\y  z^mq^nJ(m,n)\eqlabel\geneJ$$ 
As already indicated, it can be evaluated from the following limiting procedure:
$$ \Jt(z;q)= \lim_{k\rw \y} \At_{k,2k}(z;q)=(-zq)_\y\lim_{k\rw \y}  
F_{k,k}(z^2;q)\eq$$
Since  $L_k=0$, we have
$$
\eqalign{
\lim_{k\rw \y} F_{k,k}(z^2;q)  
&= \lim_{k\rw \y}\sum_{m_1,\cdots,m_{k-1}=0}^\y {
q^{N_1^2+\cdots+ N_{k-1}^2}\; z^{2N} \over (q)_{m_1}\cdots (q)_{m_{k-1}}}\cr
&= \lim_{k\rw \y}\sum_{m_1,\cdots,m_{k-1}=0}^\y {
q^{N_1^2+\cdots+ N_{k-1}^2}\; z^{2N} \over (q)_{m_1}\cdots (q)_{m_{k-1}} (z^2q)_{m_{k-1}}}\cr
& = \lim_{k\rw \y} {1\over (z^2q)_\y}\cr &
= {1\over (z^2q)_\y}\cr}
\eqlabel\andeed$$
In  the second equality we
use the fact that when $k\rw \y$, the only integer contributing to the sum over $m_{k-1}$ is precisely $m_{k-1}=0$. 
Indeed, the numerator contains the factor $q^{(k-1)m_{k-1}^2}$ which formally vanishes (with $q<1$) in the large
$k$ limit for any non-zero value of $m_{k-1}$. This allows us to introduce a judicious factor $(z^2q)_{m_{k-1}}$ in the
denominator without affecting the sum. The next equality follows from [\Andrr] (cf. theorem 2 with $M\rw
\y$). We have thus
$$ 
\Jt(z;q) = {(-zq)_\y \over (z^2q)_\y}
\eqlabel\resusu$$
Expanded in powers of $z$, this  can be rewritten as 
$$ \Jt(z;q)= \sum_{i,j=0}^\y
{q^{i(i+1)/2+j} z^{i+2j}\over  (q)_i(q)_j }\eq$$
If we define the function $\Jt_m(q)$ as
$$ \Jt(z;q)= \sum_{m=0}^\y
z^m\Jt_m(q)\eq$$
we have thus
$$\Jt_m(q) = \sum_{j=0}^{[m/2]}
{q^{(m-2j)(m-2j+1)/2+j}\over  (q)_{(m-2j)}(q)_j }\eq$$
Note that the sum could be extended to infinity by setting $1/(q)_n=0$ for $n<0$. As an example, consider the first few
terms of $\Jt_4$:\foot{This is to be compared with the
generating functions of  partitions of length 4:
$$\eqalign{
p_4(q) &= q^4 + q^5  + 2 q^6 + 3 q^7  + 5 q^8  + 6 q^9  + 9 q^{10}   + 11 q^{11}    + 15
q^{12}  \cr &  \; + 18 q^{13}    + 23 q^{14}    + 27 q^{15}    + 34 q^{16}    + 39 q^{17}    + 47
q^{18}    + 54 q^{19}+\cdots\cr}$$} 
$$\eqalign{
\Jt_4(q)&= q^2+q^3+3q^4+4q^5+7q^6+9q^7+13q^8+16q^9+22q^{10}
 +
26q^{11}\cr &+33q^{12}+39q^{13}+48q^{14}+55q^{15}+66q^{16}+75q^{17}+88q^{18}+99q^{19}+\cdots\cr}\eq$$
For instance, the nine unrestricted jagged partitions of weight 7 are: 
$$
\{(5101)\;,(4201)\;,(4111)\;,(3301)\;,(3211)\;,(2311)\;,(3121)\;,(2212)\;,(2221)\}\eq$$
(remember that these are not subject to any exclusion).

\subsec{The $q$ form of $V_r$ and bosonic characters}

This result implies that the Verma module character can be expressed as a $q$ sum, in the form
$$
\eqalign{
V_p &= \sum_{m=0}^\y \sum_{m'=0}^m \sum_{s_1,s_2=0}^\y q^{s_1} J(m+p,s_1)q^{s_2}J(m',s_2)= \sum_{m=0}^\y
\sum_{m'=0}^m J_{m+r}(q)J_{m'}(q)\cr &= \sum_{m=0}^\y \sum_{m'=0}^m\sum_{i,j=0}^\y
{q^{\frac12[(m+p-2j)(m+p-2j+1)+(m'-2i)(m'-2i+1)]+i+j}\over
(q)_{m+p-2j}\, (q)_{m'-2i}\,
(q)_i\, (q)_j}\cr}\eq$$ As already stressed, this formula holds for all values of $p$. We
stress that even thought the relative charge $r$ of the irreducible modules of interest is
supposed to be positive, some subtracted Verma modules have negative charge as (\cabo) and
(\caboo) clearly indicate. Note that the number of states in a graded parafermionic Verma module
increases very rapidly, as  illustrated in the following examples:
$$\eqalign{
V_{-3}&=
       1 + 4 q + 11 q^2   + 29 q^3  + 65 q^4  + 141 q^5
 + 287 q^6  + 564 q^7  + 1065 q^8 + 1961 q^9  + \cdots \cr
V_{-2}&=1 + 4 q + 10 q^2  + 26 q^3  + 59 q^4  + 128 q^5  + 260 q^6 + 513 q^7  + 969 q^8
+ 1787 q^9  + \cdots \cr
V_{-1}&=1 + 3 q + 9 q^2  + 22 q^3  + 52 q^4  + 111 q^5 + 230 q^6  + 451 q^7  + 861 q^8
+ 1587 q^9  +  \cdots \cr
V_{0}&=
1 + 2 q + 7 q^2  + 18 q^3  + 43 q^4  + 93 q^5  + 195 q^6  + 385 q^7  + 741 q^8
+ 1374 q^9  +  \cdots \cr
V_{1}&=  2 q + 5 q^2  + 14 q^3  + 33 q^4  + 75 q^5  + 157 q^6  + 319 q^7  + 615 q^8  +
1158 q^9  + \cdots \cr
V_{2}&= q + 3 q^2  + 10 q^3  + 24 q^4  + 57 q^5 + 122 q^6  + 253 q^7  + 495 q^8  + 945
 q^9  + \cdots \cr
V_{3}&= 2 q^2  + 6 q^3  + 17 q^4  + 40 q^5  + 92 q^6  + 191 q^7  + 387 q^8  + 745 q^9  +
\cdots \cr}\eqlabel\exavr$$
These examples also show that the leading power of $V_{p\leq 0}$ is always $q^0$ while that of $V_{p>0}$ is
$q^{[(p+1)/2]}$ as already stated.

We have now  all the ingredients for evaluating the graded bosonic characters. The results have
been compared to the fermionic formula (\andeee) for various values of
$k,\, q,\, r$ up to
$q^{20}$ typically. Note that the bosonic character becomes rapidly very involved as we go deeper into the module as
the above sample expansions (\exavr) indicate. Therefore, testing up to this level is already a very non trivial check. A
detailed illustration  of the bosonic formula (for $q=0,\, r=0,\,k=2$)  is worked out in appendix C.

Let us conclude this section with a remark. A peculiar feature of the modules $V_{r>0}$ with $r$ odd is that
there are always two terms at the leading order (cf. $V_1$ and $V_3$ in (\exavr)). These are associated to the two states
$\B_{-1}(\A_{-1})^{(r-1)/2}| \vp_\q\R$ and
$\B_0\B_{-1}(\A_{-1})^{(r-1)/2} \B_0^\da| \vp_\q\R$. In the irreducible module, one of these `top'  states could be
canceled by  a singular-vector descendent. This is certainly the case when $q=0$ since $\B_0^\da| \vp_0\R$ is
singular. Moreover, when $r=2(k+1-\q)-1$, the second state $(\A_{-1})^{(r+1)/2}\B_0^\da| \vp_\q\R$, which is
proportional to $\B_0^\da (\A_{-1})^{(r+1)/2}| \vp_\q\R$, is singular. However, in all other cases, namely $\q>0$ and
$r$ odd and such that $0<r<2(k+1-\q)-1$, two states remain at the top grade. This can also be confirmed in the fermionic
basis. For instance, with $k=3, \, q=2, \, r=1$, we verify that the descendent
$\B_{-1}\B_{-2}\B_{-1}\B_{-1}\B_{-1}\B_{0}\B_{-1}|\vp_2\R$ has the same dimension as the highest-weight state
$\B_{-1}|\vp_2\R$ .

\newsec{Conclusion}

In this work we have presented the fermionic characters of the $\osp(1,2)_k/\uh(1)$ parafermionic theory by summing up
rather directly over all the states of a given charge in the quasi-particle Hilbert space. The corresponding bosonic
characters have also been displayed. We have verified the equivalence between both expressions by comparing their direct
$q$-expansions up to reasonable order. It is of course an interesting problem to try to establish this equivalence in a
direct way and thus to all orders in
$q$.

On the other hand, the bosonic characters presented here are not exactly under the form of string functions (for the
usual parafermionic models based on $\su(2)_k/\uh(1)$, compare the analogous bosonic characters of [\ref{P. Jacob and P.
Mathieu, Nucl. Phys. {\bf B587} (2000) 514.}] with the corresponding string functions [\ref{V. Kac and D. Peterson, Adv.
Math. {\bf 53} (1984) 125}] and [\ref{D.
Nemeschansky, Nucl. Phys. {\bf B363}  (1989) 665, J. Distler and Z.  Qiu,  Nucl.Phys. {\bf B336} (1990) 533.}].)  To show
that these are different forms of the same basic expression, as expected, is also a technical issue that needs to be
addressed.

\appendix{A}{Counting unconstrained jagged partitions  from $J$-type recurrence relations}

In this appendix, we present two alternative derivations of the generating function $\Jt(z;q)$ counting the number of
unconstrained jagged partitions. The first one will illustrate another type of recurrence relations (i.e., different from
those of section 3.1) that can be found for jagged partitions. These recurrences could be of interest in other
contexts. Moreover, this method can be lifted to the analysis of $k$-constrained jagged partitions, as shown in
appendix B.  The second method uses the correspondence between unconstrained jagged partitions and $\Z_3$
constrained ordinary partitions.

\subsec{Recurrence relations for unconstrained jagged partitions }

 We first derive a simple recurrence
relation for the $J(m,n)$, the set of jagged partitions of length $m$ and weight $n$.
We introduce the following pictural representation:
$$J(m,n): \qquad (\cdots 010101^+) \qquad (m\;{\rm entries)}\eq$$ 
Here the symbol $+$ indicates that we can build up the jagged partition on the ground state up to the position of  the
$+$. Therefore, in the set
$(\cdots 0101^+01)$ the filling process stops at the penultimate `1'; this set is thus equivalent to that of  all
jagged partitions that have at least one pair of ``01'' at the end. With the number of entries always  understood to
be $m$, it should
 be clear that $(\cdots 0101^+01)$ is a pictural representation of the set $J(m-2,n-1)$:
$$J(m-2,n-1): \qquad (\cdots 0101^+01) \eq$$
In other  words, the subset of $J(m,n)$ of jagged partitions ending with a ``01'' is in a one-to-one
correspondence with the set $J(m-2,n-1)$.

In this notation, our recurrence relation reads:
$$(\cdots 010101^+) =( \cdots 0101^+01) +(\cdots 10101^+1) -(\cdots 101^+011) +(\cdots 121212^+) \eqlabel\partu$$
Indeed, a jagged partition can end with a ``01'' - which accounts  for the first term on the right hand side, or simply
a ``1'', which corresponds to the second term or even a larger integer, which is the last term. All these sets
with specified tail are exclusive. However, there is a potential problem with the second term because it could
contain terms that are not genuine jagged partitions: these are those particular terms built on $(\cdots 10101^+1)$
that terminates with ``011''.  These must thus be eliminated and this is the reason for the subtracted third term. 
Rewritten in terms of $J(m,n)$, (\partu) becomes:
$$J(m,n)= J(m-2,n-1)+J(m-1,n-1)-J(m-3,n-2)+J(m,n-m)\eqlabel\jform$$
The retranscription should be clear except maybe for the last term.  The set $(\cdots 121212^+)$ can be characterized
by the following property: subtracting from these elements the partition $(\cdots 111111)$ (reducing the weight by
$m$ but without affecting the length)  make them elements of
$J(m,n-m)$.

At the level of the  generating function (\geneJ),
the relation (\jform) becomes
$$\Jt(z;q)=zq\Jt(z;q)+ z^2q\Jt(z;q)-z^3q^2 \Jt(z;q) +\Jt(zq;q)\eqlabel\jformgf$$
whose solution is
$$\Jt(z;q)= {1\over (1-zq)(1-z^2q)}\Jt(zq;q)\eq$$
Iterating this result, assuming that $q<1$, so that $\lim_{n\rw\y}zq^n=0$ and using $\Jt(0,q)=1$, we get
$$\eqalign{
\Jt(z;q) &= \prod_{n=1}^\y{1\over (1-zq^n)(1-z^2q^{2n-1})}={1 \over (zq)_\y (z^2q;q^2)_\y } \cr
& = {(z^2q^2;q^2)_\y \over (zq)_\y (z^2q)_\y} =
{(-zq)_\y \over  (z^2q)_\y}\cr}
\eq$$
in agreement with the result (\resusu) of  section 5.3.

 \subsec{A `pseudo tableaux' transcription of the recurrence relations}

As an addendum to the previous subsection, we point out
 a natural description of (\jform) in terms of `pseudo tableaux'.

To make this point clear, let us derive
the analogous recurrence of ordinary partitions
$p(m,n)$ of $n$ into $m$ parts in the language of tableaux. Such partitions are in correspondence with Young
tableaux of $n$ boxes distributed among $m$ rows whose length does not increase when read from top to bottom. The
set of tableaux of $m$ rows can be decomposed into the union of those tableaux with precisely one box on the last
row and those with more that one box on this
$m$-th row. The former set is characterized by the fact that removing the single box on the $m$-th row leaves a
tableau with $m-1$ rows (and of course a total of $n-1$ boxes). Similarly, the tableaux with two boxes or more on
the $m$-th row are characterized by the fact that they still have $m$ rows if we take out the first column (reducing
$n$ to
$n-m$). That gives:
$$p(m,n)=p(m-1,n-1)+p(m,n-m)\eq$$

Consider now the case of jagged partitions. They correspond to `pseudo Young tableaux', that is tableaux for which the
non-increasing condition on the row lengths is modified: the length can increase by one box from top to bottom but cannot
increase if we compare lengths by skipping one row each time. The number of pseudo tableaux with $m$ rows can again
be decomposed into two sets: those with a single box in the $m$-th row and those with more that one boxes in the
$m$-th row. The second set is equivalent to $J(m,n-m)$. The first set however is split into those tableaux with
no boxes in the $(m-1)$-th row, described by the set $J(m-2,n-1)$, and those with one or more boxes in the $(m-1)$th
row, given by $J(m-1,n-1)$. From the latter set, we need to exclude all tableaux of $m-3$ rows built over a tail with
no boxes in the
$(m-2)$-th row and one box in the next two rows.  This reproduces thus (\jform).

\subsec{A bijection between unconstrained jagged partitions and $\Z_3$
constrained  partitions} 

Still another method for  deriving the generating function $\Jt(z;q)$ counting the number of
unconstrained jagged partitions is based on
a bijection between jagged partitions $(n_1,\cdots,n_m)$ and 
usual partitions
$(\la_1,\cdots,\la_m)$ (with $\la_i\geq \la_{i+1}$) subject to a standard
$\Z_3$ exclusion principle that enforces the difference condition:
$$\la_i\geq \la_{i+2}+2\eq$$
This is the $k=3$
specialization of the more general condition: $\la_i\geq \la_{i+k-1}+2$.
 The bijection is the
following:
$$\la_i=n_i+m-i\eq$$
In this way, the `ground state' $(\cdots 01010101)$ is mapped to the staircase with doubled-length steps $(\cdots
77553311)$. 

The generating function for these $\Z_3$ constrained partitions is of course a special case of the
Andrews' result [\Andrr] already quoted in section 3.2:
$$F_{3,3}(z;q)= \sum_{m_1,m_2=0}^\y {q^{(m_1+m_2)^2+m_2^2}\,  z^{m_1+2m_2}\over (q)_{m_1}(q)_{m_2}}\eq$$
The coefficient of $z^{N}$  gives the total number of $\Z_3$ constrained partitions of length $N$. To make contact with
the generating function of jagged partitions, we need to take out the contribution of the staircase $(N-1\cdots 543210)$
of weight $N(N-1)/2$: this is achieved by introducing the term $q^{-N(N-1)/2}$ with $N=m_1+2m_2$ within the sum (now
denoted
${\tilde F}_{3,3}(z;q)$):
$$\eqalign{
{\tilde F}_{3,3}(z;q) &= \sum_{m_1,m_2=0}^\y { q^{-(m_1+2m_2)(m_1+2m_2-1)/2} 
q^{(m_1+m_2)^2+m_2^2}\,  z^{m_1+2m_2}\over
(q)_{m_1}(q)_{m_2}}\cr &=  \sum_{m_1,m_2=0}^\y {q^{m_1(m_1+1)/2+m_2} \, z^{m_1+2m_2}\over (q)_{m_1}(q)_{m_2}}\cr
&=  {(-zq)_\y\over (z^2q)_\y}= \Jt(z;q)\cr}
\eq$$

 \appendix{B}{Counting $k=2$ constrained jagged partitions using $J$-type recurrence relations}

Let us now turn to the counting of constrained jagged partitions using the method of the previous appendix.
Unfortunately, these recurrence relations, in contradistinction with those presented in section 3.1, are not
recurrence on the boundary parameter
$i$. In addition, they depend strongly upon
$k$, i.e., their structure becomes rapidly very complicated as $k$ increases. Here we consider only the case $k=2$. 
These relations will provide another check of the results of section 3, albeit for a very special case. 

In terms of the symbolic notation introduced in appendix A, we can write
$$
\eqalign{
(\cdots101^+) &= (\cdots212^+) + (\cdots212^+1) + (\cdots212^+01) \cr
& + [(\cdots212^+11)-(\cdots323^+1211)- (\cdots323^+21211)]\cr
& +(\cdots212^+101) + [(\cdots212^+111)- (\cdots323^+12111)- (\cdots323^+212111)]\cr}\eq$$
In this expression, we have decomposed the partitions according to the form of their tail, paying due care to the
exclusion conditions (which are $n_i\geq n_{i+3}+1$ or $n_i=n_{i+3}$ and $n_{i+1}=n_{i+2}+2$) that enforce various
subtractions (the corrections of a given set of partitions are delineated by square brackets). These relations can be
written in terms of the
$k=2$ constrained jagged partitions of length
$m$ and weight $n$, denoted $\p(m,n)$. They become
$$
\eqalign{
\p(m,n)&=\p(m,n-m)+\p(m-1,n-m)+\p(m-2,n-m+1)\cr
& +\p(m-2,n-m)-\p(m-4,n-2m+3)-\p(m-5,n-2m+3)
\cr &+ \p(m-3,n-m+1) +\p(m-3,n-m)\cr &-\p(m-5,n-2m+4)-\p(m-6,n-2m+4)\cr}\eq$$
Lifting this relation at the level of generating functions,
with
$$\P(z;q)= \sum_{n,m=0}^\y q^nz^m\p(m,n)\eq$$
we get
$$
\eqalign{
\P(z;q) &= (1+zq+z^2q+z^2q^2+z^3q^2+z^3q^3)\P(zq;q)\cr
&\qquad -z^4q^5(1+zq^2+zq+z^2q^3)\P(zq^2;q)\cr
&= (1+zq)(1+z^2q+z^2q^2)\P(zq;q)-z^4q^5(1+zq)(1+zq^2)\P(zq^2;q)\cr}\eq$$
Let us now make contact with the results obtained in section 3.2. Since in the previous counting we did not took into
account any boundary condition, the generating function so obtained has to be equivalent to $A_{2,4}(z;q)$. Indeed, the
above relation is exactly the same as (\aqaq), which establishes the equivalence $\P(z;q)= A_{2,4}(z;q)$.

\appendix{C}{The zero-relative charge vacuum $k=2$ bosonic character}

Let us detail the construction of the bosonic character for the special case $k=2,\,q=r=0$. In the following tables, we
present the singular vectors, obtained by acting with the operator displayed in the column SV on all those above which
preceed it (in order) and this string is understood to act on the vacuum $|\vp_\q\R$. We also give the charge of the
resulting singular vector, the operator that maps it to the $r=0$ module and the dimension of the corresponding string
(which is thus the lowest dimensional folded
 string).  For type-I singular vectors, we have:
 $$ \eqalignSS{
& \qquad & {\rm SV}\qquad & \#\A\qquad &\#\B^\da\qquad &{\rm charge}\qquad &{\rm folding}\qquad &{\rm dim}\cr
&{\rm I(i)_0}: \qquad & \B_0^\da         &0  &1  &-1  &\B_{-1}        &1\cr
&{\rm I(ii)_0}: \qquad & \A_{-1}^4       &4  &1  &\phantom{-}7   &(\B_{0}^\da)^7  &4\cr
&{\rm I(i)_1}: \qquad & (\B_0^\da)^{15}  &4 &16  &-8  &\A_{-1}^4         &8\cr
&{\rm I(ii)_1}: \qquad & \A_{-1}^{11}    &15 &16 &\phantom{-}14  &(\B_{0}^\da)^{14}  					&15\cr
&{\rm I(i)_2}: \qquad & (\B_0^\da)^{29}  &15 &45 &-15 &\B_{-1}\A_{-1}^7  					&23\cr
&{\rm I(ii)_2}: \qquad & \A_{-1}^{18} &33 &45 &\phantom{-}21& (\B_{0}^\da)^{21}    				&33\cr}\eq$$
while for type-II, the data are
$$ \eqalignSS{
& \qquad & {\rm SV}\qquad & \#\A\qquad &\#\B^\da\qquad & {\rm charge}\qquad &{\rm folding}\qquad &{\rm dim}\cr
&{\rm II(i)_0}: \qquad & \A_{-1}^3          &3  &0   &\phantom{-}6  &(\B_{0}^\da)^6       &3\cr
&{\rm II(ii)_0}: \qquad & (\B_0^\da)^{13}   &3  &13   &-7           & \B_{-1}\A_{-1}^3  &7\cr
&{\rm II(i)_1}: \qquad & \A_{-1}^{10}     &13 &13  &\phantom{-} 13  &(\B_{0}^\da)^{13}         &13\cr
&{\rm II(ii)_1}: \qquad &  (\B_0^\da)^{27}  &13 &40  &-14           &  \A_{-1}^7					&20\cr
&{\rm II(i)_2}: \qquad &  \A_{-1}^{17}  &30 &40   &\phantom{-}20    & 	(\B_{0}^\da)^{20}				&30\cr
&{\rm II(ii)_2}: \qquad & (\B_0^\da)^{41}   &30 &81  &-21           & \B_{-1}\A_{-1}^{10}     				&41\cr}\eq$$

Up to the factor $q^{1/28}$, the character is obtained from $V_0$ minus the contribution of the (i)-type vectors plus
that of the (ii)-type ones. For the above table, we thus read off 
$$
\eqalign{ 
M_{0,0}&=  V_0-V_1+q^4V_{-7}-q^4V_8+q^{15}V_{-14}-q^{15}V_{15}+q^{33}V_{-21}+\cdots\cr
&\;\;-q^3V_{-6}+q^3V_{7}-q^{13}V_{-13}+q^{13}V_{14}-q^{30}V_{-20}+q^{30}V_{21}+\cdots\cr}\eqlabel\maneuf$$
With $V_0$ and $V_1$ already given in (\exavr) and the expansions:
$$
\eqalign{
V_{-7} &=1+4q+12q^2+32q^3+75q^4+165q^5+341q^6+676q^7+1287q^8+2381q^9+\cdots\cr
V_{-6}& =1+4q+12q^2+32q^3+74q^4+162q^5+334q^6+660q^7+1253q^8+2314q^9+\cdots\cr
V_{7}& =2q^4+6q^5+18q^6+44q^7+103q^8+218q^9+\cdots\cr
V_{8}& =q^4+3q^5+11q^6+28q^7+69q^8+151q^9+\cdots\cr}\eq$$
substituted into (\maneuf), we find that 
$$M_{0,0}=1+2q^2+3q^3+7q^4+10q^5+18q^6+26q^7+44q^8+62q^9+\cdots\eq$$
This agrees with the expansion of the fermionic character formula (\andeee) for $k=2$ and $q=r=0$.

\vskip0.3cm
\centerline{\bf Acknowledgment}

PJ and PM would like to thank IPAM and the organizers of the CFT 2001 semester, where this work was started, 
for their hospitality. We also thank A. Schilling for her collaboration at the initial phase of this work  and  for
useful discussions. This work is supported by NSERC.

\vskip0.3cm

\centerline{\bf REFERENCES}

\immediate\closeout\refs \vskip 0.5cm
  \message{References}\input references
\vfill\eject

\end